\input{aipcheck}

\documentclass[
    ,final            
  ]
  {aipproc}

\layoutstyle{6x9}

\makeatletter
\def\spig{\kern.15em\raise1.2ex\hbox{$|$}\kern-.48em\to} 
\def\anti#1{\mathpalette{\@anti}{#1}#1}
\def\@anti#1#2{\sbox0{$#1#2$}
  \makebox[0pt][l]
    {$#1\kern.30\ht0\overline{\kern-.35\ht0\phantom{#2}\kern-.1ex}$}}
\makeatother
\begin{document}

\title{Light quark and charm interplay in the Dalitz-plot analysis
of hadronic decays in FOCUS}

\author{S.Malvezzi}{
  address={I.N.F.N, Milan, Via Celoria 16-
        20133 Milan, Italy}
}

\begin{abstract}
The potentiality of interpreting the $D$-meson decay-dynamics has revealed
itself to be strongly dependent on our understanding of the light-meson sector.
The statistics collected by FOCUS is already at a level that manifests
parametrization problems for scalar particles. In this paper the first
application of the $K$-\emph{matrix} approach in the charm sector is
illustrated and preliminary results on the $D^+$ and $D_s$ decays to three
pions are shown.
\end{abstract}

\maketitle


\section{Introduction}
Charm-meson decay-dynamics has been studied extensively over the last decade.
Dalitz plot analysis has indeed revealed itself a powerful tool for
investigating the effects of resonant substructures, interference patterns and
final-state interactions in the charm sector. The isobar formalism
traditionally applied consists of a sum of relativistic Breit--Wigner
resonances properly modulated by form factors (Blatt--Weisskopf barrier factors
are generally used) and multiplied by an angular term assuring angular-momentum
conservation. The potentiality to investigate charm dynamics has revealed
itself to be strongly connected with a knowledge of the light-meson sector. In
particular, the need to model the scalar particles populating our charm-meson
Dalitz plots has led us to question the validity of the BW approximation for
the description of a resonance. Resonances are associated with poles of the
$S$-matrix in the complex energy plane. It is the position of the pole that
provides the fundamental, model-independent, process-independent resonance
parameters.

The fitting of data with simple Breit--Wigner formul{\ae} corresponds to the most
elementary type of extrapolation from the physical region to an
unphysical-sheet pole. In the case of a narrow, isolated resonance, there is a
close connection between the position of the pole on the unphysical sheet and
the peak we observe in experiments at real values of the energy. However, when
a resonance is broad and overlaps with other resonances, this connection is
lost. The BW parameters measured on the real axis (mass and width) can be
connected to the pole-positions in the complex energy plane only through models
of analytic continuation. A formalism for studying overlapping and many channel
resonances was proposed long ago and is based on the
$K$-\emph{matrix}\cite{wigner,chung} parametrization. This formalism,
originating in the context of two-body scattering, can be generalized to cover
the case of resonance production in more complex reactions, with the
assumptions that the two-body system in the final state is isolated and that
the two particles do not simultaneously interact with the rest of the final
state in the production process\cite{aitch}. Its implementation allows us to
include the positions of the poles in the complex plane directly in our
analysis, embedding in our amplitudes the results from spectroscopy
experiments\cite{penn1,anisar1}.

 \section{The fit formalism}

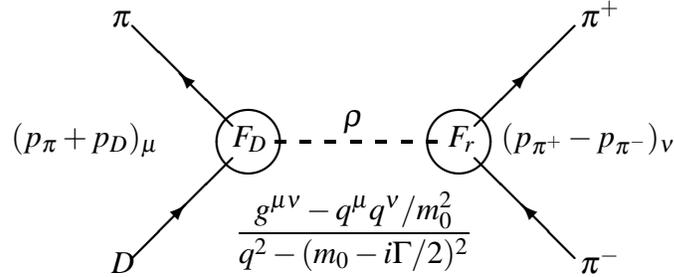
\begin{figure}[htb]
  \setlength{\unitlength}{0.7mm}
  \centering
  \begin{picture}(110,50)(0,0)
    \thicklines
    \put(30,25){\circle{12}}
    \put(27,24){$F_D$}
    \put(70,25){\circle{12}}
    \put(68,24){$F_r$}
    \put(-15,24){$(p_\pi+p_D)_\mu$}
    \put(78,24){$(p_{\pi^+}-p_{\pi^-})_\nu$}
    \put(48,28){$\rho$}
    \put(25,-2){\makebox(50,20){$\displaystyle\frac
                                 {g^{\mu\nu}-q^\mu q^\nu/m_0^2}
                                 {q^2-(m_0-i\Gamma/2)^2}$}}
    \put(04,47){$\pi$}
    \put(93,47){$\pi^+$}
    \put(04,00){$D$}
    \put(93,00){$\pi^-$}
    \put(27,28){\vector(-1,1){10}}
    \put(73,28){\vector(1,1){10}}
    \put(18,37){\line(-1,1){10}}
    \put(82,37){\line(1,1){10}}
    \put(27,22){\line(-1,-1){10}}
    \put(73,22){\line(1,-1){10}}
    \put(08,03){\vector(1,1){10}}
    \put(92,03){\vector(-1,1){10}}
    \multiput(35,25)(4.8,0){7}{\line(1,0){2.0}}
  \end{picture}
  \caption{The $D^+\to\pi\pi\pi$ decay diagram.}
  \label{fig:fey}
\end{figure}

The formalism traditionally applied to three-body charm decays relies on the
so-called isobar model. A resonant amplitude for a quasi-two-body channel, of
the type
\begin{displaymath}
  \begin{array}{cl}
    D \to &  r + c     \\[-0.5ex]
          & \spig a + b
  \, ,
  \end{array}
\end{displaymath}
is interpreted \emph{{\`a} la} Feynman. For the decay $D\to\pi\pi\pi$ of
Fig.~\ref{fig:fey},
a $D\to\pi$ current with form factor $F_D$ interacts with a di-pion current
with form factor $F_r$ through an unstable propagator with an imaginary width
contribution in the propagator mass. Each resonant decay function is thus,
\begin{equation}
  A =
  F_DF_r
  \times
  |\bar c|^J |\bar a|^J P_J(\cos\Theta^r_{ac})
  \times
  BW(m_{ab})
  \,
  \label{twobody}
\end{equation}
i.e., the product of two vertex form factors (Blatt--Weisskopf
momentum-dependent factors), a Legendre polynomial of order $J$ representing
the angular decay wave function, and a relativistic Breit--Wigner (BW). In this
approach, already applied in the previous analyses of the same channels
\cite{e687_dpds,e791_dpds}, the total amplitude (Eq.~\ref{totamp}) is assumed
to consist of a constant term describing the direct non-resonant three-body
decay and a sum of functions (Eq.~\ref{twobody}) representing intermediate
two-body resonances.

\begin{equation}
  A(D) = a_0 e^{i\delta_0} + \sum_i a_i e^{i\delta_i} A_i \, ,
   \label{totamp}
\end{equation}

Additional care has to be applied to describe the $f_0(980)$ resonance because
of the opening of the $K\anti K$ channel near its pole mass. A Flatt\'e
coupled-channel parametrization \cite{e687_dpds,flatte} is generally adopted.

\subsection{The \boldmath{$K$}-\emph{matrix} formalism}

For a well-defined wave of specific isospin and spin \emph{IJ}, characterized
by narrow and isolated resonances the propagator is, as anticipated, of the
simple BW form. In contrast, when the specific wave \emph{IJ} is characterized
by large and heavily overlapping resonances, just as the scalars, the
propagation is no longer dominated by a single resonance, but is the result of
complicated interplay among the various resonances. In this case, it can be
demonstrated on very general grounds that the propagator may be written in the
context of the $K$-\emph{matrix} approach as
\begin{equation}
(I -iK \cdot \rho)^{-1}
 \label{eq_prop}
\end{equation}
where \emph{K} is the matrix for the scattering of particle $a$ and $b$ and
$\rho$ is the phase-space matrix.

While the need for a $K$-\emph{matrix} parametrization, or in general for a
more accurate description than the isobar model, is questionable for the vector
and tensor amplitudes, since the resonances are relatively narrow and well
isolated, this parametrization is unavoidable for the correct treatment of the
scalar amplitudes. Indeed the $\pi\pi$ scalar resonances are large and overlap
each other in such a way that it is impossible to single out the effect of any
one of them on the real axis.

In order to write down the propagator, we need the scattering  matrix. To
perform a meaningful fit to our three-pion data, we thus need a full
description of the scalar resonances in the relevant energy range, updated to
the most recent measurements in this sector. To our knowledge, the only
self-consistent description of $s$-wave isoscalar scattering is that given in
the $K$-\emph{matrix} representation by Anisovich and Sarantsev in
\cite{anisar1} through a global fit of all the available scattering data from
the $\pi\pi$ threshold up to 1900\,MeV.

To introduce the $K$-\emph{matrix} formalism used in this analysis, we must
recall some general, non-trivial, considerations. The production of an
isoscalar, $s$-wave state with an accompanying pion, involves, in the energy
region relevant to this analysis, five channels, namely $1=\pi\pi$,
$2=K\anti{K}$, $3=\eta\eta$, $4=\eta\eta'$ and $5=\;$multi-meson states (mainly
four-pion states at $\sqrt{s}<1.6$\,GeV). The production amplitude in the
particular channel $(00^{++})_i\pi$ can be written as
\begin{equation}
  F_i = (I-iK\rho)_{ij}^{-1}P_j
  \label{eq_f}
\end{equation}
where $I$ is the identity matrix, $K$ is the $K$-\emph{matrix} describing the
isoscalar $s$-wave scattering process, $\rho$ is the phase-space matrix for the
five channels, and $P$ is the `initial' production vector into the five
channels. In this picture, the production process is viewed as consisting of an
initial preparation of several states, which then propagate via the term
$(I-iK\rho)^{-1}$ into the final state. In particular, the three-pion final
state can be fed by an initial formation of $(\pi\pi)\pi$, $(K\anti K)\pi$,
etc. The particular process we are dealing with, $(\pi^+\pi^-)\pi^+$, is then
described by the amplitude $F_1$.

In this paper, the $K$-\emph{matrix} used is that of \cite{anisar1}:
\begin{equation}
 K_{ij}^{00}(s) =
  \left\{
    \sum_\alpha \frac{g^{(\alpha)}_i g^{(\alpha)}_j}{m^2_{\alpha}-s}
 +  f^\mathrm{scatt}_{ij}\frac{1\,\mathrm{GeV}^2 - s_0^\mathrm{scatt}}{s-s_0^\mathrm{scatt}}
  \right\}
 \times \frac{s-s_A/2m^2_{\pi}}{(s-s_{A0})(1-s_{A0})}.
  \label{eq_sarantsev}
\end{equation}
The factor $g^{(\alpha)}_i$ is the coupling constant of the
$K$-\emph{matrix}\,\footnote{Note that the $K$-\emph{matrix} masses and widths
do not need to be identical to those of the $T$-\emph{matrix} poles in the
complex energy plane.} pole $\alpha$ to meson channel $i$; the parameters
$f^\mathrm{scatt}_{ij}$ and $s_0^\mathrm{scatt}$ describe a smooth part of the
$K$-\emph{matrix} elements; the factor
$\frac{s-s_A/2m^2_{\pi}}{(s-s_{A0})(1-s_{A0})}$ suppresses a false kinematical
singularity in the physical region near the $\pi\pi$ threshold (Adler zero).
The parameter values used in this analysis are listed in
Table~\ref{table_sara}.


\begin{table}[htb]
\begin{tabular}{cccccc}
\hline \tablehead{1}{r}{b}{K-matrix \\mass poles}&
\tablehead{1}{r}{b}{\boldmath{$g_{\pi\pi}$}} &
\tablehead{1}{r}{b}{\boldmath{$g_{K \anti K}$}} & \tablehead{1}{r}{b}
{\boldmath{$g_{4 \pi}$}} & \tablehead{1}{r}{b}{\boldmath{$g_{\eta \eta}$}} &
\tablehead{1}{r}{b}
{\boldmath{$g_{\eta \eta'}$}}\\
\hline
0.65100  & 0.24844& -0.52523 & 0.00000& -0.38878& -0.36397 \\
1.20720  & 0.91779&  0.55427 & 0.00000&  0.38705&  0.29448 \\
1.56122  & 0.37024&  0.23591 & 0.62605&  0.18409&  0.18923 \\
1.21257  & 0.34501&  0.39642 & 0.97644&  0.19746&  0.00357 \\
1.81746  & 0.15770& -0.17915 &-0.90100& -0.00931&  0.20689 \\
\hline $s_0^\mathrm{scatt}$ & $f^\mathrm{scatt}_{11}$ & $f^\mathrm{scatt}_{12}$
& $f^\mathrm{scatt}_{13}$ & $f^\mathrm{scatt}_{14}$
& $f^\mathrm{scatt}_{15}$ \\
-3.30564 & 0.26681 & 0.16583 & -0.19840 & 0.32808 & 0.31193\\
 \hline
$s_A$ & $s_{A0}$ & & & & \\
1.0 & -0.2 & & & & \\
\hline
\end{tabular}
\caption{$K$-\emph{matrix} parameters.} \label{table_sara}
\end{table}
These K-matrix values correspond to a $T$-matrix description of five poles
whose masses and half-widths, in GeV, are (1.019,\,0.038), (1.306,\,0.167),
(1.470,\,0.960), (1.489,\,0.058) and (1.749,\,0.165); the analysis in
\cite{anisar1} does not require the $\sigma$. The decay amplitude for the $D$
meson into three-pion final state, where $\pi^+\pi^-$ are in a
$(IJ^{PC}=00^{++}$)-wave, is thus written as
\begin{eqnarray}
  A(D\to (\pi^+\pi^-)_{00^{++}}\pi^+)= F_1 =
  (I-iK\rho)^{-1}_{1j} \qquad&\qquad\nonumber \\
  \times \left\{
    \sum_\alpha \frac{\beta_{\alpha}g_{j}^{(\alpha)}}{m^2_{\alpha}-m^2}
  + f_{1j}^\mathrm{prod}\frac{1\,\mathrm{GeV}^2 - s_0^\mathrm{prod}}{s-s_0^\mathrm{prod}}
  \right\}
  \times \frac{s-s_A/2m^2_{\pi}}{(s-s_{A0})(1-s_{A0})}.
  \label{eq_kmat}
\end{eqnarray}
where $\beta_{\alpha}$ is the coupling to the pole $\alpha$ in the `initial'
production process, $f_{1j}^\mathrm{prod}$ and $s_0^\mathrm{prod}$ are the
$P$-\emph{vector} background parameters. In the end, the complete decay
amplitude of the $D$ meson into three-pion final state is:
\begin{equation}
  A(D) = a_0 e^{i\delta_0} + \sum_i a_i e^{i\delta_i} A_i + F_1
  \label{totampK}
\end{equation}
where the index $i$ now runs only over the vector and tensor resonances, which
can be safely treated as simple Breit--Wigner's (see Eq.~\ref{twobody}). In the
fit to our data, the $K$-\emph{matrix} parameters are fixed to the values of
Table~\ref{table_sara}, which consistently reproduce measured s-wave isoscalar
scattering. The free parameters are those peculiar to the $P$-\emph{vector},
i.e.\ $\beta_{\alpha}$, $f^\mathrm{prod}_{1j}$ and $s_0^\mathrm{prod}$, and
those in the remaining isobar part of the amplitude, $a_i$ and $\delta_i$.

\section{Three-pion preliminary results}

\begin{figure}[]
\includegraphics[height=.3\textheight]{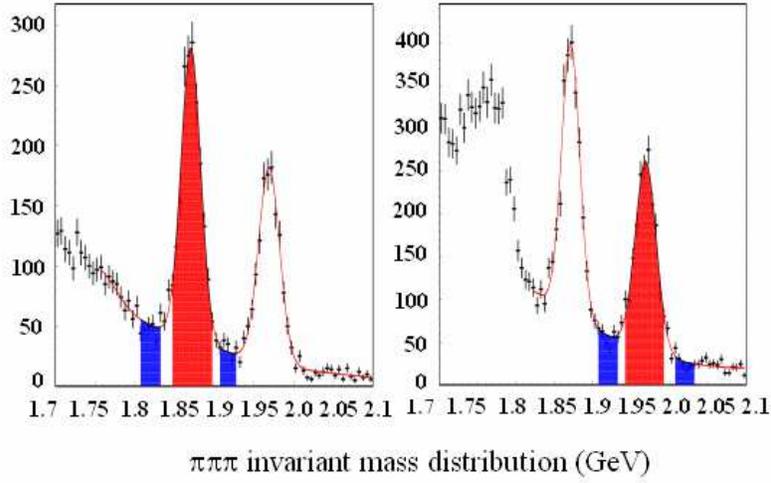}
  \caption{$D^+$ and $D_s$ signals and sideband regions.}
  \label{mass}
\end{figure}

The three-pion selected samples (Fig.~\ref{mass}) consist of $1527\pm51$ and
$1475\pm50$ events for the $D^+$ and $D_s$ respectively. The Dalitz plot
analyses are performed on yields within $\pm2\sigma$ of the fitted mass value.

\begin{figure}[]
\includegraphics[height=.3\textheight]{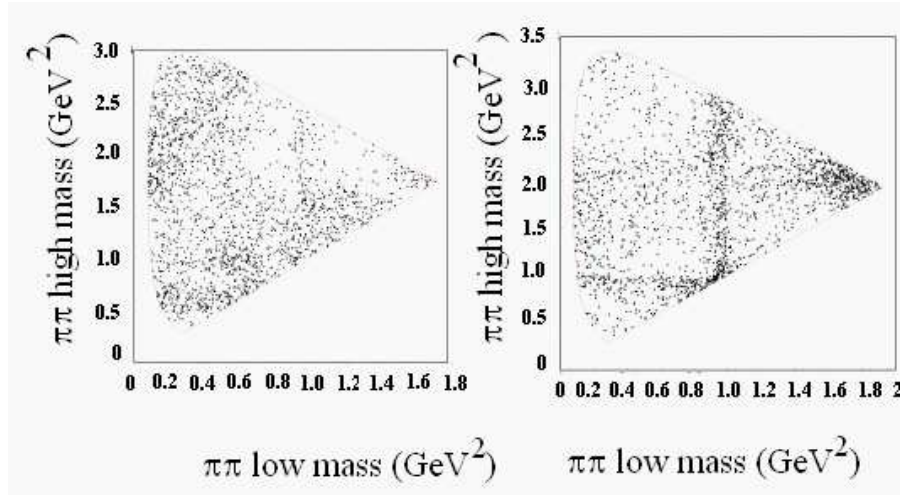}
  \caption{Three-pion $D^+$ and $D_s$ Dalitz plots.}
  \label{dalitz}
\end{figure}

\subsection{\boldmath{$D_s \to \pi \pi \pi$}}

The $D_s$ Dalitz plot (Fig.~\ref{dalitz}) is clearly structured. Indeed, two
$f_0(980)$ bands in the low and high projection at 1000\,MeV, a band in the
1500\,MeV high-projection region and an accumulation of events in the corner of
the folded plot are easily recognized.

\begin{table}[h]
\begin{tabular}{ccc}
\hline
resonance & fit fraction(\%) & phase (deg) \\
\hline
NR             & 25.5 $\pm$ 4.6 & 246.5 $\pm$ 4.7  \\
$f_2(1275)$    &  9.8 $\pm$ 1.3 & 140.2 $\pm$ 9.2  \\
$f_0(980)$     & 94.4 $\pm$ 3.8 &  0(fixed)        \\
$S_0(1475$     & 17.4 $\pm$ 3.1 & 249.7 $\pm$ 6.4  \\
$\rho^0(1450)$ &  4.1 $\pm$ 1.0 & 187.3 $\pm$ 15.3 \\
\hline
\end{tabular}
\caption{$D_s$ fit fractions and phases with the isobar model.}
 \label{table1}
\end{table}

\begin{figure}[]
\includegraphics[height=.3\textheight]{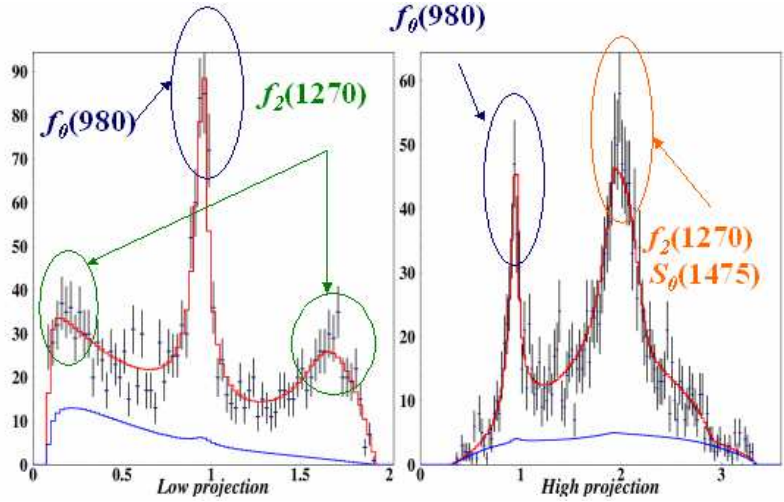}
 \caption{$D_s$ Dalitz plot projections and isobar fit results}
\label{ds_proj}
\end{figure}

When the isobar model is applied, satisfactory fits can be obtained letting
only the $f_0(980)$ and $f_0(1500)$ parameters float freely. We obtain $m_0=975
\pm 10$\,MeV, $\Gamma_{\pi \pi}=90 \pm 30$\,MeV and $\Gamma_{KK}=35 \pm
15$\,MeV for the $f_0(980)$ and $m_0 = 1475 \pm 10 $\,MeV and $\Gamma_0 = 112
\pm 24$\,MeV for the $f_0(1500)$. The fit C.L., evaluated with a $\chi^2$
estimator over a Dalitz plot with bin size adaptively chosen according to the
statistics is 11.5\%. The fit results\footnote{A fit fraction is conventionally
defined as the ratio between the intensity for the single amplitude integrated
over the Dalitz plot and that of the total amplitude with all the modes and
interferences present.} are reported in Table~\ref{table1}.

\begin{table}[]
\begin{tabular}{cccc}
\hline
resonance & fit fraction(\%) & phase (deg)  \\
\hline
$s-wave$       &  87.8 $\pm$ 1.7  & 0(fixed)            \\
$f_2(1275)$    &  11.5 $\pm$ 1.2  &  120.3 $\pm$ 5.4    \\
$\rho^0(1450)$ &   5.4 $\pm$ 1.1  &  198.6 $\pm$ 9.3    \\
\hline
\end{tabular}
\caption{ $D_s$ fit fractions and phases with the $K$-\emph{matrix} model.}
 \label{table3}
\end{table}

All the established resonances decaying to $\pi^+\pi^-$ with a sizeable
branching ratio are first considered in the fit. The set of contributions
reported in Table~\ref{table1} refers to fit coefficients of more than
$2\sigma$ statistical significance. A fit with the $f_0(1500)$ parameters fixed
to the PDG values has been also performed; the C.L. is 1\% and the non-resonant
fit fraction increases to about 40\% (to be compared with the value in
Table~\ref{table1}), indicating how the flat non-resonant component, not having
particular signatures, absorbs the residual parametrization problems. A large
systematic effect should thus be associated with this channel decay fraction
weakening the potentiality to establish a reliable estimate of the annihilation
processes. The projection fit results are shown in Fig.~\ref{ds_proj}. It is
interesting to note that the sum of the fit fractions is about 150\%, pointing
to a large, rather suspicious, interference effect among the contributions.

\begin{figure}[]
\includegraphics[height=.3\textheight]{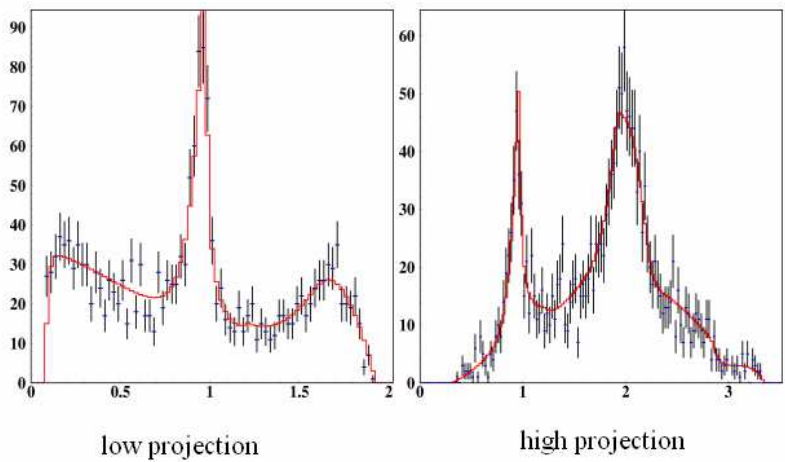}
\caption{$D_s$ Dalitz-plot projections with $K$-\emph{matrix} fit results.}
\label{ds_proj_Kmatrix}
\end{figure}

One of the main advantages of the $K$-\emph{matrix} approach is the
straight-forward prescription for constraining the amplitudes to respect the
two-body unitarity and the universality of the poles in the complex energy
plane. The resonance parameters are fixed to the reference paper
\cite{anisar1}. The results are quoted in Table~\ref{table3}. The contribution
of the whole $s$-wave in the $D_s$ decay into three pions is computed in a
single fit fraction. The fit C.L. is 3.2\%. It is interesting to note here that
the sum of the fit fractions is very close to 100\%. The fractions of
$f_2(1275)$ and $\rho^0(1450)$ are quite consistent in the two approaches; the
main difference concerns the scalar sector. A good fit to our $D_s\to3\pi$
signal can be obtained even without the non-resonant contribution. The
$K$-\emph{matrix} fit results are shown on the Dalitz projections in
Fig.~\ref{ds_proj_Kmatrix}.

\subsection{\boldmath{$D^+ \to \pi \pi \pi$}}

\begin{figure}[]
\includegraphics[height=.3\textheight]{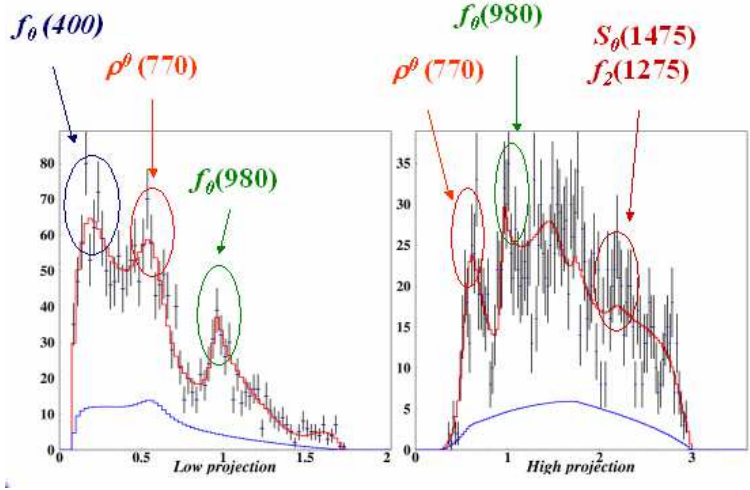}
\caption{$D^+$ Dalitz-plot projections and the isobar model fit results.}
 \label{dp_proj}
\end{figure}

The $D^+\to\pi^+\pi^-\pi^+$ channel shows an excess of events at low $\pi\pi$
mass, (Fig.~\ref{dp_proj}), which cannot be explained with simple BW's
corresponding to well-established resonances. A fit performed with a free
parameter single-channel Breit--Wigner returns  a mass $m=443\pm27$ and width
$\Gamma=443\pm80$. The decay fraction and phases are reported in
Table~\ref{table2}. The $f_0(980)$ parameters are fixed to the values found in
the $D_s$ channel, where its signal is statistically more robust. The fit C.L.
is about 10\%; when the low-energy BW, $\sigma$, is removed the C.L. drops to
$10^{-8}$.

\begin{table}[h]
\begin{tabular}{cccc}
\hline
resonance & fit fraction(\%) & phase (deg)  \\
\hline
NR            &  9.8 $\pm$ 4.3 &   0(fixed)              \\
$f_2(1275)$   & 12.3 $\pm$ 2.1  & -213.3 $\pm$ 17.7       \\
$f_0(980) $   &  6.7 $\pm$ 1.5  & -145.9 $\pm$ 17.7       \\
$\rho(770)$   & 32.8 $\pm$ 3.8 &   62.9 $\pm$ 16.8       \\
$S_0(1475)$   &  1.8 $\pm$ 1.2  &  242.3 $\pm$ 25.8       \\
$f_0(400) $   & 18.9 $\pm$ 5.3  & - 96.9 $\pm$ 30.7       \\
\hline
\end{tabular}
\caption{$D^+$ fit fractions and phases with the isobar model.}
 \label{table2}
\end{table}

\begin{figure}[h]
\includegraphics[height=.3\textheight]{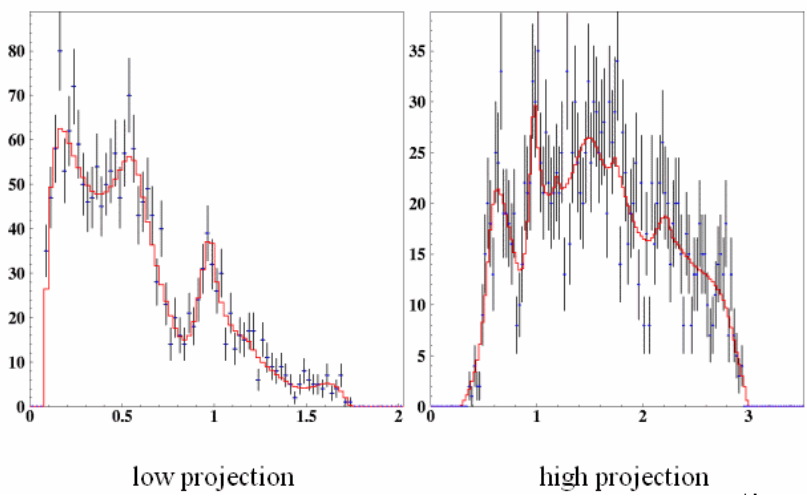}
\caption{$D^+$ Dalitz-plot projections with the $K$-\emph{matrix} formalism.}
\label{dp_proj_Kmatrix}
\end{figure}

When the $K$-\emph{matrix} approach is applied to the $D^+$, it is found that
the $P$-\emph{vector} degrees of freedom are able to reproduce the features of
the low-energy $\pi\pi$ distribution with a fit C.L. of about 3.5\%. The
$K$-\emph{matrix} fit results on the $D^+$ Dalitz projections are shown in
Fig.~\ref{dp_proj_Kmatrix}; the fit fractions and phases are reported in
Table~\ref{table4}.

\begin{table}[]
\begin{tabular}{ccc}
\hline
resonance & fit fraction(\%) & phase (deg) \\
\hline
$s-wave$       &  66.5 $\pm$ 4.2  & 101.8    $\pm$ 22.5    \\
$f_2(1270)$    &  11.4 $\pm$ 1.4  & 247.0   $\pm$ 9.0     \\
$\rho^0(770)$  &  21.2 $\pm$ 4.4  & 0(fixed)                \\
\hline
\end{tabular}
\caption{$D^+$ fit fractions and phases with the $K$-\emph{matrix} model.}
 \label{table4}
\end{table}

\section{Conclusions}
The $K$-\emph{matrix} approach has been applied for the first time to the
three-pion Dalitz-plot analysis in FOCUS. The results are extremely encouraging
since the same parametrization of two-body resonances coming from light-quark
experiments also works for charm decays; this result was not obvious
beforehand. In the $D^+\to\pi\pi\pi$ channel a good C.L. fit is obtained even
without the $\sigma$, suggesting that a lot of work has still to be done before
solving the $\sigma$ puzzle. A global fit including the charm data and the
analysis of the $KK\pi$ channel could add important experimental information.
The method explained here will find its full application to forthcoming
excellent statistics of the charm experiments.




\bibliographystyle{aipproc}   

\bibliography{sample}

\begin{thebibliography}{99}





\bibitem{wigner} E.P. Wigner, Phys.~Rev~70 (1946) 15.

\bibitem{chung} S.U. Chung \emph{et al.}, Ann.~Physik~4 (1995) 404.

\bibitem{aitch} I.J.R. Aitchison, Nucl.~Phys.~A187 (1972) 417.

\bibitem{penn1} K.L. Au, D. Morgan, M.R. Pennington, Phys.~Rev.~D35 (1987) 1633;
 M.R. Pennington hep-ph/9905241.

\bibitem{anisar1} V.V. Anisovich, A.V. Sarantsev, Eur.~Phys.~J.~A16 (2003) 229.

\bibitem{e687_dpds} P.L. Frabetti \emph{et al.}, Phys.~Lett.~B407 (1997) 79.

\bibitem{e791_dpds} E.M. Aitala \emph{et al.}, Phys.~Rev.~Lett.~86 (2001) 765;
Phys.~Rev.~Lett.~86 (2001) 770.









\bibitem{flatte} S.M. Flatt\'e, Phys.~Lett.~B63 (1976) 228.

\end{thebibliography}

\IfFileExists{\jobname.bbl}{}
 {\typeout{}
  \typeout{******************************************}
  \typeout{** Please run "bibtex \jobname" to optain}
  \typeout{** the bibliography and then re-run LaTeX}
  \typeout{** twice to fix the references!}
  \typeout{******************************************}
  \typeout{}
 }

\end{document}